\documentclass[10pt]{article}

\usepackage{amsmath}
\usepackage{amssymb}
\usepackage{graphicx}
\usepackage{cite}
\usepackage{hyperref}
\usepackage{longtable}

\usepackage{color}
\usepackage[labelfont=bf,labelsep=period,justification=raggedright]{caption}
\makeatletter
\renewcommand{\@biblabel}[1]{\quad#1.}
\makeatother

\date{}
\newcommand{\hJ}{\hat{J}}
\newcommand{\Lk}{\mathcal{L}}
\newcommand{\SKRR}{\mathsf{SKRR}}
\newcommand{\SK}{\mathsf{SK}}
\newcommand{\RR}{\mathsf{RR}}
\newcommand{\R}{\textsuperscript{\textregistered}}
\newcommand{\MATLAB}{MATLAB\R}
\newcommand{\Meff}{M_{\mathrm{eff}}}
\newcommand{\Pind}{P^{\mathrm{ind}}}
\newcommand{\Pdir}{P^{\mathrm{dir}}}

\begin{document}

\begin{flushleft}
{\Large
\textbf{Fast and accurate multivariate Gaussian modeling of protein families: Predicting residue contacts and protein-interaction partners}
}
\\
Carlo Baldassi$^{1,2,\dagger}$,
Marco Zamparo$^{1,2,\dagger}$,
Christoph Feinauer$^1$,
Andrea Procaccini$^2$,
Riccardo Zecchina$^{1,2}$,
Martin Weigt$^{3,4}$,
Andrea Pagnani$^{1,2,\ast}$
\\
$^1$ Department of Applied Science and Technology and Center for Computational Sciences, Politecnico di Torino,  Torino, Italy
\\
$^2$ Human Genetics Foundation-Torino, Torino, Italy
\\
$^3$ Sorbonne Universit\'es, Universit\'e Pierre et Marie Curie Paris 06, UMR 7238, Computational and Quantitative Biology, Paris, France
\\
$^4$ Centre National de la Recherche Scientifique, UMR 7238, Computational and Quantitative Biology, Paris, France
\\
$\ast$ E-mail: andrea.pagnani@polito.it
\\
$\dagger$ These authors contributed equally to this work

\end{flushleft}

\section*{Abstract}
In the course of evolution, proteins show a remarkable conservation of
their three-dimensional structure and their biological function,
leading to strong evolutionary constraints on the sequence variability
between homologous proteins. Our method aims at extracting such
constraints from rapidly accumulating sequence data, and thereby at
inferring protein structure and function from sequence information
alone. Recently, global statistical inference methods
(e.g.~direct-coupling analysis, sparse inverse covariance estimation)
have achieved a breakthrough towards this aim, and their predictions
have been successfully implemented into tertiary and quaternary
protein structure prediction methods. However, due to the discrete
nature of the underlying variable (amino-acids), exact inference
requires exponential time in the protein length, and efficient
approximations are needed for practical applicability. Here we propose
a very efficient multivariate Gaussian modeling approach as a variant
of direct-coupling analysis: the discrete amino-acid variables are
replaced by continuous Gaussian random variables. The resulting
statistical inference problem is efficiently and exactly solvable. We
show that the quality of inference is comparable or superior to the one achieved
by mean-field approximations to inference with discrete variables, as
done by direct-coupling analysis. This is true for \emph{(i)}~the
prediction of residue-residue contacts in proteins, and
\emph{(ii)}~the identification of protein-protein interaction partner
in bacterial signal transduction. An implementation of our
multivariate Gaussian approach is available at the website
\url{http://areeweb.polito.it/ricerca/cmp/code}.

\section*{Introduction}

One of the most important challenges in modern computational biology
is to exploit the wealth of sequence data, accumulating thanks to
modern sequencing technology, to extract information and to reach an
understanding of complex biological processes. A particular example is
the inference of conserved structural and functional properties of
proteins from the empirically observed variability of amino-acid sequences in
homologous protein families, e.g.~via the inference of signals of
co-evolution between residues, which may be distant along the
sequence, but in contact in the folded protein;
cf.~\cite{Altschuh1987, Gobel1994, Neher1994,Shindyalov1994,
Lockless1999,Fodor2004} for a selection of classical works and \cite{valencia2013}
for a review over recent developments. In the last 5 years, a
strong renewed interest in residue co-evolution has been
emerging: a number of global statistical inference
approaches~\cite{weigt, burger, Morcos2011, Balakrishnan,
Jones, Dijk, MonassonPlosCompBio2013, Aurell2013, Kamisetty2013} have led to a highly increased
precision in predicting residue contacts from sequence information
alone.  Furthermore, co-evolutionary analysis was found to provide
valuable insight on specificity and partner prediction in
protein-protein interaction~\cite{nimwegen, proca} in bacterial signal
transduction.

Key to this recent progress are \emph{global statistical inference}
approaches, like \emph{direct-coupling analysis} (DCA)~\cite{weigt,
Morcos2011} and \emph{sparse inverse covariance estimation}
(PSICOV)~\cite{Jones}, and the GREMLIN algorithm based on \emph{pseudo-likelihood
maximization}~\cite{Balakrishnan, Kamisetty2013}. DCA is based on the maximum-entropy (MaxEnt)
principle~\cite{JaynesA, JaynesB} which naturally leads to statistical
models of protein families in terms of so-called Potts models or
Markov random fields. Proposed initially more than a
decade ago~\cite{Lapedes, lapedes02}, it was not until very recently that the
first successful MaxEnt approaches to the study of co-evolution were
published~\cite{weigt, mora}. The main idea behind such global
inference techniques is the following: correlations between the
amino-acids occurring in two positions in a protein family,
i.e.~between two columns in the corresponding multiple-sequence
alignment (MSA), may result not only from direct co-evolutionary
couplings. They may also be generated by a whole network of such
couplings.  More precisely, if a position $i$ is
coupled to a position $j$, and $j$ is coupled to $k$, then $i$ and $k$
will also show some correlation even if they are not coupled. The aim
of global methods is to disentangle such direct and indirect effects,
and to infer the network of direct co-evolutionary couplings starting from
the empirically observed correlations.

In this context, we focus on two different biological problems: the
inference of residue-residue contacts and the prediction of
interaction partners.

The inference of residue-residue contacts from large MSAs of
homologous proteins~\cite{weigt, burger, Morcos2011, Balakrishnan,
  Jones, Dijk, MonassonPlosCompBio2013, Aurell2013, Kamisetty2013} is
an important challenge in structural biology. Inferred contacts have
been shown to be sufficient to guide the assembly of complexes between
proteins of known (or homology modeled) monomer
structure~\cite{schug09, Dago2012}, and to predict the fold of single
proteins~\cite{noiplosone2011, taylorA, nugent2012, sulkowska2012,
  taylor, hopf2012}, including highlights like large trans-membrane
proteins~\cite{hopf2012, nugent2012}. In~\cite{Dago2012}, the
predicted structure of the auto-phosphorylation complex of a bacterial
histidine sensor kinase has been used to repair a non-functional
chimeric protein by rationally designed mutagenesis; this structure is
also, to the best of our knowledge, the first case of a prediction,
which has subsequently been confirmed by experimental X-ray
structures~\cite{VicK2013, Diensthuber2013}. The possibility to guide
tertiary and quaternary protein structure prediction is an important
finding, in light of the experimental effort needed for generating
high-resolution structures.

The second problem, concerning molecular determinants of interaction
specificity of proteins and the identification of interaction
partners~\cite{nimwegen, proca}, is a central problem in systems
biology.  In both cited papers, bacterial two-component signal
transduction systems (TCS) were chosen, which constitute a major way
by which bacteria sense their environment, and react to it~\cite{TCS}.
TCS consist of two proteins, a histidine sensor kinase (SK) and a
response regulator protein (RR): the SK senses an extracellular signal, and
activates a RR by phosphorylation; the RR typically acts as a
transcription factor, thus triggering a transcriptional response to
the external signal. The same (homologous) phosphotransfer mechanism
is used for several signaling pathways in each bacterium; thus, to
produce the correct cellular response to an external signal,
interactions have to be highly specific inside each pathway:
crosstalk between pathways has to be avoided~\cite{Jim, Laub,
Hendrik}. This evolutionary pressure can be detected by
co-evolutionary analysis~\cite{nimwegen, proca}.  Results are
interesting: statistical couplings inferred by DCA reflect physical
interaction mechanisms, with the strongest signal coming from charged
amino-acids. They are able to predict interacting SK/RR pairs for
so-called orphan proteins (SK and RR proteins without an obvious
interaction partner), and the predictions compared favorably to most
available experimental results, including the prediction of 7 (out of
8 known) interaction partners of orphan signaling proteins in
\emph{Caulobacter crescentus}~\cite{proca}.

In the present study, we describe an alternative approach to
co-evolutionary analysis, based on a multivariate Gaussian modeling of
the underlying MSA\@. It can be understood as an approximation to the MaxEnt
Potts model in which \emph{(i)}~the discreteness constraint is released,
i.e.~continuous values are allowed for variables representing amino-acids,
\emph{(ii)}~a Gaussian interaction model is assumed, and \emph{(iii)}~a prior
distribution is introduced to compensate for the under-sampling of the data.
This simplification allows to explicitly determine the model parameters from
empirically observed residue correlations.  The approach shares many
similarities with~\cite{Jones}, in which a multivariate Gaussian model is also
assumed, and with the mean-field approximation to the discrete DCA
model~\cite{Morcos2011}, but the simpler structure of the probability
distribution makes the model analytically tractable, and allows for an
efficient implementation, while still having a prediction accuracy comparable
or superior to that of the aforementioned models (see the Results section). The
model is briefly described in the next section, and in greater detail in the
Materials and Methods section.

A fast, parallel implementation of the multivariate Gaussian modeling approach
is provided on \url{http://areeweb.polito.it/ricerca/cmp/code} in two different
versions, a \MATLAB~\cite{matlab} one and a Julia~\cite{julia} one.

\section*{Gaussian modeling of multiple sequence alignments}

This section briefly outlines the prediction procedure coming from our
proposed model, and highlights its main distinctive features with
respect to other similar methods. A full presentation can be found in
the Materials and Methods section, and additional details in the
Supporting Informations Section.

The input data to our model is the MSA for a large protein-domain family,
consisting of $M$ aligned homologous protein sequences of length $L$. Sequence
alignments are formed by the $Q=20$ different amino-acids, and may contain
alignment gaps.

As in~\cite{Jones}, we consider a multivariate Gaussian model in which each
variable represents one of the $Q$ possible amino-acids at a given site, and
aim in principle at maximizing the likelihood of the resulting probability
distribution given the empirically observed data (in particular, given the
observed mean and correlation values, computed according to a reweighting
procedure devised to compensate for the sampling bias). Doing so would yield
the parameters for the most probable model which produced the observed data,
which in turn would provide a synthetic description of the underlying
statistical properties of the protein family under investigation.
Unfortunately, however, this is typically infeasible, due to under-sampling of
the sequence space. A possible approach to overcome this problem, used
e.g.~in~\cite{Jones}, is to introduce a sparsity constraint, in order to reduce
the number of degrees of freedom of the model. Here, instead, we propose a
Bayesian approach, in which a suitable prior is introduced, and the parameter
estimation is then performed over the posterior distribution.

A convenient choice for the prior is the normal-inverse-Wishart (NIW), which,
being the conjugate prior of the multivariate Gaussian distribution, provides a
NIW posterior. Thus, within this choice, the posterior simply is a
data-dependent re-parametrization of the prior: as a result, the problem is
analytically tractable, and the computation of relevant quantities can be
implemented efficiently. Furthermore, by choosing the parameters for the prior
to be as uninformative as possible (i.e.~corresponding to uniformly distributed
samples), we obtain an expression for the posterior which, interestingly, can
be reconciled with the pseudo-count correction of~\cite{Morcos2011}: in the
Gaussian framework, the pseudo-count parameter has a natural interpretation as
the weight attributed to the prior.

We then estimate the parameters of the model as averages on the posterior
distribution, which have a simple analytical expression and can be computed
efficiently (in practical terms, the computation amounts to the inversion of a
$LQ\times LQ$ matrix).  On one hand, this yields an estimate of the strengths
of direct interactions between the residues of the alignments, which can be
used to predict protein contacts.  On the other hand, this allows to build
joint models of interacting proteins, which can be used to score candidate
interaction partners, simply by computing their likelihood - which can be done
very efficiently on a Gaussian model.

The contact prediction between residues relies on the model's inferred
interaction strengths (i.e.~couplings), which are represented by
$Q\times Q$ matrices; in order to rank all possible interactions, we
need to compute a single score out of each such matrix. As mentioned
above, these matrices are numerically identical to those obtained in
the mean-field approximation of the discrete (Potts) DCA model. We
tested two scoring methods: the so-called direct information (DI),
introduced in~\cite{weigt}, and the Frobenius norm (FN) as computed
in~\cite{Aurell2013}. The DI is a measure of the mutual information
induced only by the direct couplings, and its expression is
model-dependent: in the Gaussian framework it can be computed
analytically (see the Supporting Information Section) and yields slightly
different results with respect to the Potts model (but with a
comparable prediction power, see the Results section). The FN, on the
other hand, does not depend on the model, and therefore some of the
results which we report here for the contact prediction problem are
applicable in the context of the Potts model as well.  In our tests,
the FN score yielded better results; however, the DI score is
gauge-invariant and has a well-defined physical interpretation, and is
therefore relevant as a way to assess the predictive power of the
model itself.

%Importantly, though, the interpretation of the coupling matrices is different
%in the Gaussian framework, and this is particularly relevant in the context of
%interaction partners prediction -- since it allows to define a log-likelihood
%score in a straightforward way -- and, potentially, for further improving the
%prediction quality.

%Our approach also shares many similarities with~\cite{Jones}, in which
%a multivariate Gaussian model is also assumed (and in fact we also use
%the exact same sequence reweighting scheme in out tests, see Methods),
%but our method for inferring the couplings, based on an \emph{a posteriori}
%estimate with a conjugate prior based on a uniform distribution, is
%significantly more computationally efficient.
%
%Prediction accuracy of residue contacts is comparable or superior to that achieved
%in~\cite{Morcos2011} or by using the PSICOV method of~\cite{Jones} with
%default settings; accuracy in pairing interaction partners is comparable to
%that achieved in~\cite{proca} (cf.~the Results section).
%
%Therefore, thanks to the simple structure of the
%probability distribution, which allows for an analytical computation of many
%relevant quantities (e.g.~likelihoods and posterior probabilities), this
%Gaussian approximation scheme turns out to be an extremely efficient
%method for reconstructing protein-protein interactions.

\section*{Results}

\subsection*{Residue-residue contact prediction}

The aim of the original DCA publication~\cite{weigt} was the
identification of inter-protein residue-residue contacts in protein
complexes, more precisely in the SK/RR complex in bacterial signal
transduction. More recently, global methods for inferring direct
co-evolution attacked the problem prediction of intra-domain contacts
for large protein domain families~\cite{burger, Morcos2011,
Balakrishnan, noiplosone2011, Jones, Dijk, MonassonPlosCompBio2013,
Aurell2013, Kamisetty2013}. Thanks to the
development of more efficient approximation techniques triggered by
the wide availability of single-domain data on databases like Pfam~\cite{pfamNew},
one can now easily undertake co-evolutionary analysis of a large
number of protein families on normal desktop computer. To give a
comparison, whereas the message-passing algorithm in~\cite{weigt} was
limited to alignments with up to about 70 columns at a time (typically
requiring some ad-hoc pre-processing of larger alignments to select
the 70 potentially most interesting columns), the subsequent
approaches easily handle MSA of proteins with up to ten times
this number of columns.

In this context, our multivariate Gaussian DCA is particularly
efficient: parameter estimation can be done explicitly in one step,
and the computation of the relevant coupling measures such as the
direct information (DI) and the log-likelihood also uses explicit
analytical formulae.  The analytical tractability of Gaussian
probability distributions results in a major advantage in algorithmic
complexity, and therefore in real running time.  In the included
implementation of the algorithm the largest alignment analyzed
(PF00078, $L=214$ residues, $M=126258$ sequences) the DI is obtained
in about 20 minutes, whereas a more typical alignment (e.g. PF00089,
$L = 219$, $M = 15894$) is analyzed in less than a minute on a normal
$@2270$ MHz Intel\R Core i5 M430 CPU on a Linux desktop.  With respect
to the computational complexity of the algorithm, the sequence
reweighting step is $\mathcal{O}\left(M^2 L\right)$ (since it requires
a computation of sequence similarity for all sequence pairs in the
MSA), while the model's parameters estimate is
$\mathcal{O}\left(L^3\right)$ (since it requires to invert a
covariance matrix whose size is proportional to $L$).

\begin{figure}[!h]
\centering
\includegraphics[width=0.5\textwidth]{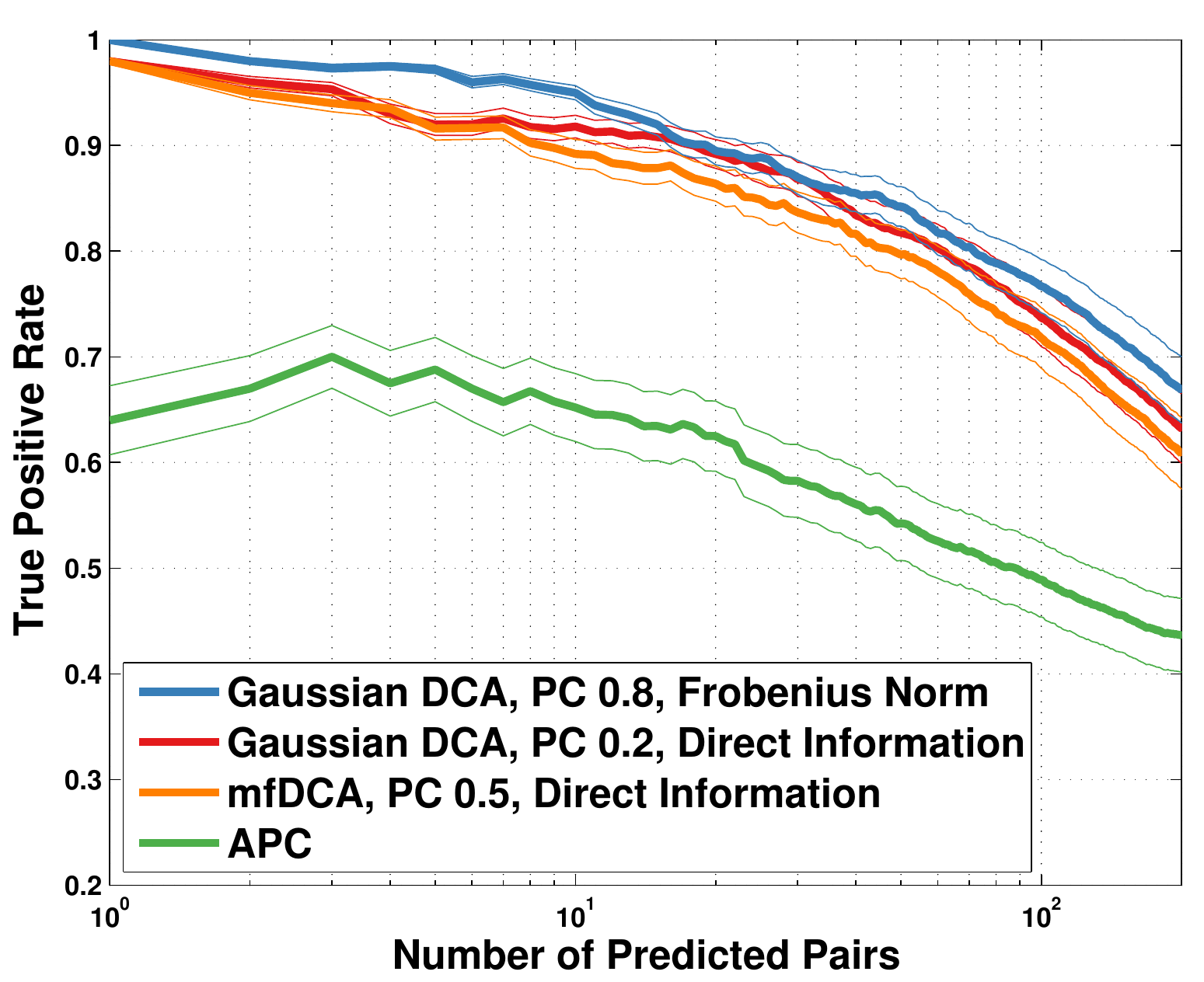}
\caption{True positive rate plotted against number of predicted pairs.
Results are shown for four different different scoring techniques:
Frobenius norm (as described in~\cite{Aurell2013}, pseudo-count set
to $0.8$, blue); Gaussian direct information (as described in the text,
APC-corrected, pseudo-count set to $0.2$, red); mean-field direct information
(as described in~\cite{Morcos2011}, pseudo-count set to $0.5$, orange) and
APC-corrected mutual information (as described in~\cite{dunn2008mutual},
green). The true positive rate is an arithmetic mean
over 50 Pfam families (see Table~\ref{table:pfam_table} for the
list); thin lines represent standard deviations.}
\label{fig:tpr}
\end{figure}

Here, we will show that this gain in running time has no detectable
cost in terms of predictive power. To this aim, we first studied the
prediction of intra-domain contacts (see Fig.~\ref{fig:tpr}). From the
Pfam database~\cite{pfamNew}, a set of 50 families was selected for
which the number of representative sequences is high enough to allow
for a meaningful statistical analysis (average length $\langle L
\rangle = 173.48$ residues, average number of sequences per alignment
$\langle M \rangle = 32660.2$), cf.~the Methods section. For each
family, 4 measures were determined: DI in mean-field approximation, DI
and Frobenius norm (FN) in the Gaussian model,
Average-product-corrected mutual information (MI) as described
in~\cite{dunn2008mutual}. As mentioned above, the FN in the Gaussian
model is the same as that computed in the mean-field approximation of
the discrete DCA model. Each measure was used to rank residue position
pairs (only pairs which are at least 5 positions apart in the chain
are considered), and high-ranking pairs are evaluated according to
their spatial proximity in exemplary protein structures. A cutoff of
8\AA~minimal distance between heavy atoms for contacts was chosen, in
agreement with~\cite{Morcos2011} and~\cite{Garbuzynskiy04}. The best
overall results are obtained with FN, as already noted
in~\cite{Aurell2013}; however, it is interesting to note that the
Gaussian DI score is comparable to, and even slightly better then the
mean-field DI score, which gives an important indication regarding the
accuracy of the underlying probabilistic model: this in turn is
relevant for subsequent analysis (see next section). Somewhat
surprisingly, we also found that the optimal overall value of the
pseudo-count parameter is strongly dependent on which scoring function
is used: we explored the whole range $\left(0,1\right)$ in steps of
$0.1$, and found that the optimum for the FN score was at $0.8$, while
for the DI score it was at $0.2$.

\begin{figure}
\centering
\includegraphics[width=0.5\textwidth]{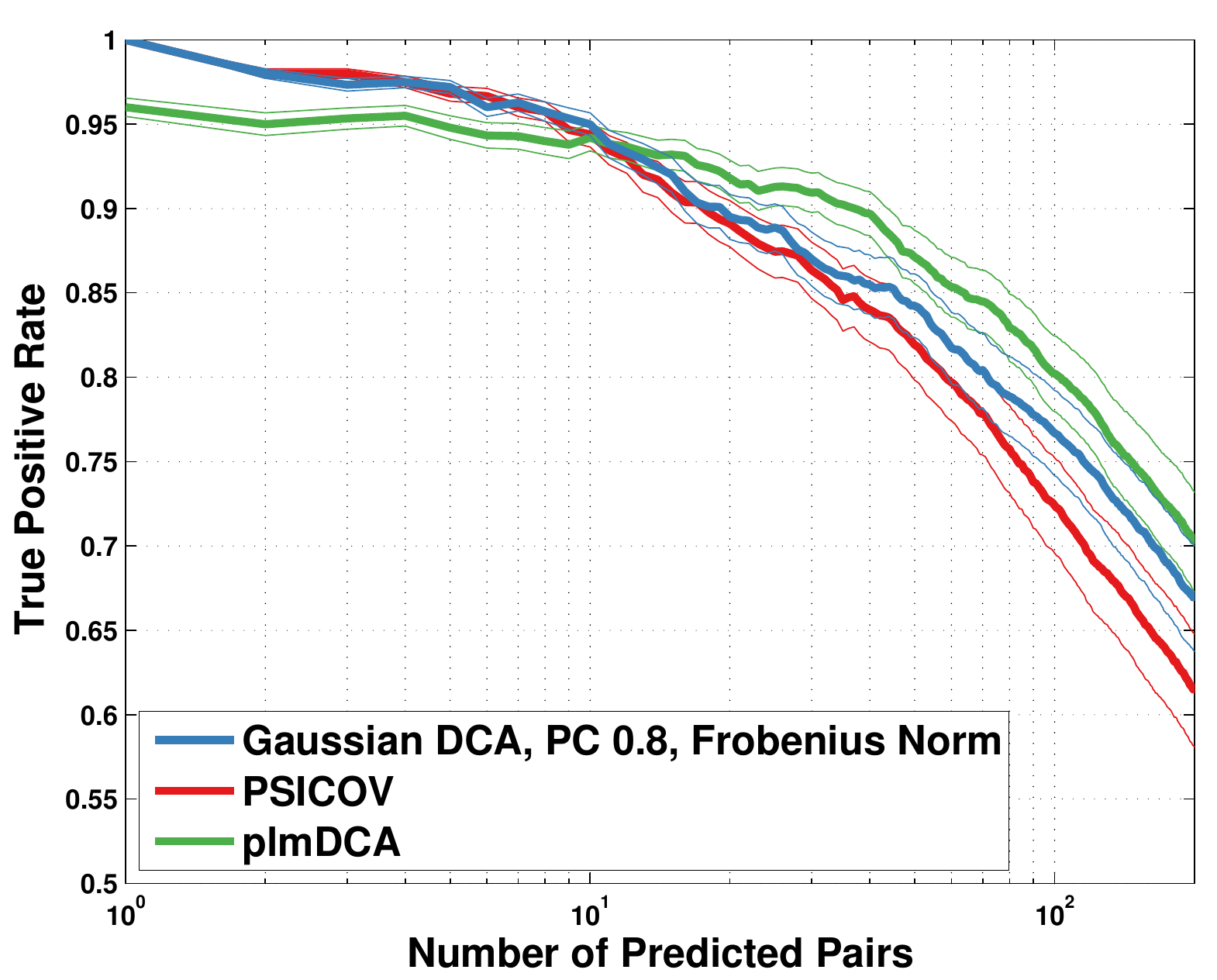}
\caption{True positive rate plotted against number of predicted pairs.
Data for plmDCA~\cite{Aurell2013} (green) and PSICOV version
1.11~\cite{Jones} (red) was obtained using the code provided by the
authors with standard parameters as found in the distributed code,
except that PSICOV was run with the \texttt{-o} flag to override
the check against insufficient effective number of sequences.
The true positive rate is an arithmetic mean over 50 Pfam families
(see Table~\ref{table:pfam_table} for the list); thin lines represent
standard deviations.}
\label{fig:psicov}
\end{figure}

\begin{table}[h]
\centering
\begin{tabular}{|c|cccc|}
\hline
 & PF00014 & PF00025 & PF00026 & PF00078 \\
 \hline
N & 53 & 175 & 317 & 214 \\
M & 4915 & 5460 & 4762 & 172360 \\
\hline
Gaussian DCA (parallel) & 0.7 & 5.3 & 16.3 & 534.8 \\
Gaussian DCA (non-parallel) & 1.7 & 12.7 & 52.1 & 3583.4 \\
PSICOV & 11.7 & 1141.9 & 5442.7 & 10965.1 \\
plmDCA & 433.2 & 6980.7 & 37364.8 & 303331.0 \\
\hline
\end{tabular}
\caption{Running times in seconds for a representative sample of
  proteins with varying length ($N$) and sequences in alignment ($M$),
  using different algorithms.  Since the Gaussian DCA code is
  parallelized, we show two series of results, one in which we used 8
  cores and one in which we forced the code to run on a single core,
  for the sake of comparing with the non-parallel code of PSICOV and
  plmDCA\@. These benchmarks were taken on a $48$-core cluster of
  $2100.130$ MHz AMD Opteron 6172 processors running Linux 3.5.0;
  PSICOV version 1.11 was used, compiled with gcc 4.7.2 at
  \texttt{-O3} optimization level; plmDCA was run with \MATLAB version
  r2011b.  Gaussian DCA timings shown are taken using the Julia
  version of the code, using Julia version 0.2.}
\label{table:running_times}
\end{table}

As a second test we ran on the same data-set a direct comparison
between our method's best score, PSICOV~\cite{Jones} and
plmDCA~\cite{Aurell2013}.  Fig.~\ref{fig:psicov} shows that our
method's performance is comparable to that of PSICOV (and even
marginally better after the first 50 inferred couplings), and that the
two methods are slightly better for the first 10 predicted contacts
(with a 100\% accuracy on the first contact).  At ten predicted
contacts, the true positive average is about 95\% for all three
methods. From ten predicted pairs on, both our method and PSICOV
perform slightly worse than plmDCA: at 100 predicted contacts, the
true positive rate is about 72\% for PSICOV, 77\% for the Gaussian
model and 80\% for plmDCA\@. A sample of running times for the three
methods and different problem sizes, reported in
Table~\ref{table:running_times}, shows that our code can be at least
an order of magnitude faster then PSICOV, and two orders of magnitude
faster then plmDCA\@. These results suggest that our method is a good
candidate for large scale problems of inference of protein contacts.

\begin{figure}
\centering
\includegraphics[width=0.48\textwidth]{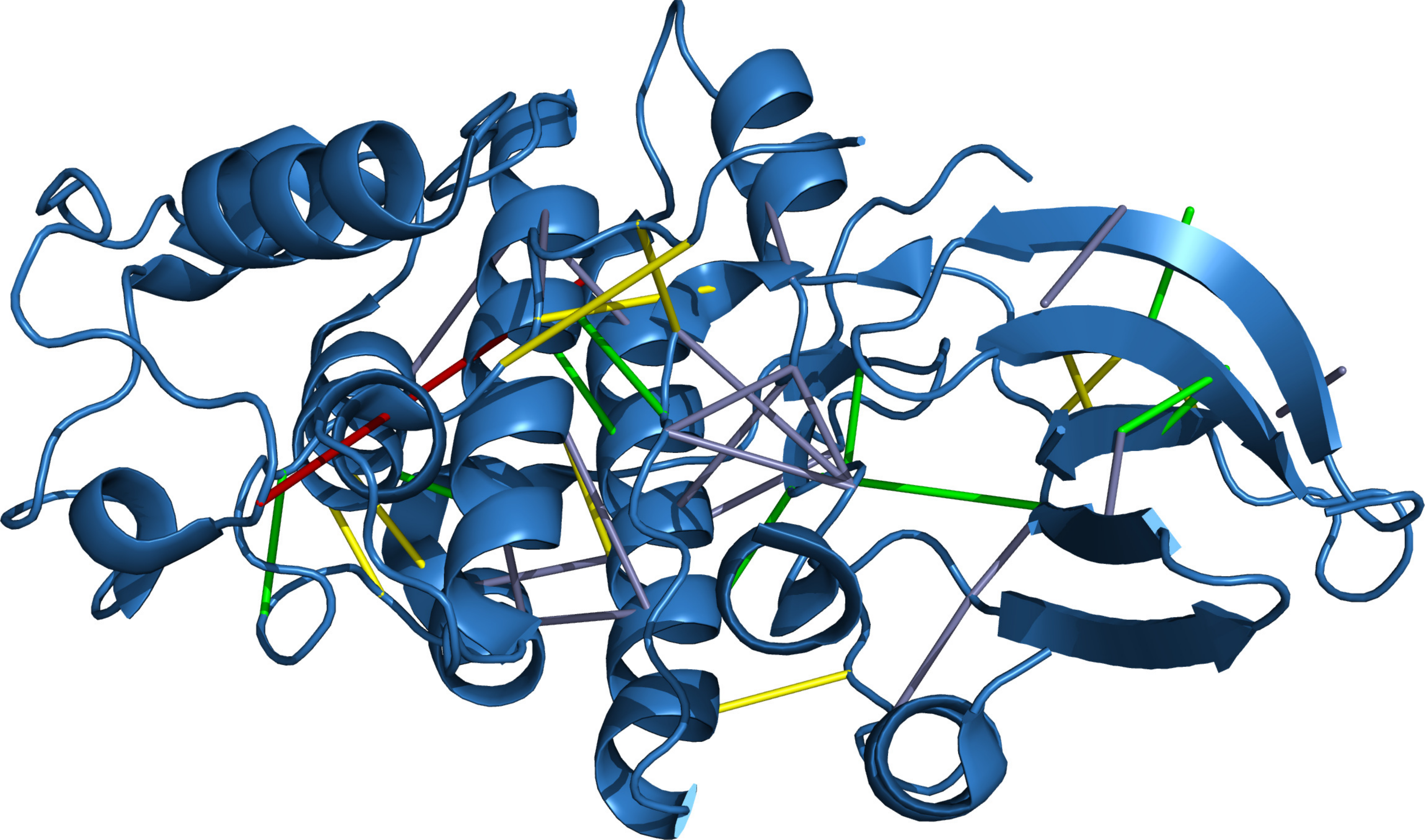}
\includegraphics[width=0.48\textwidth]{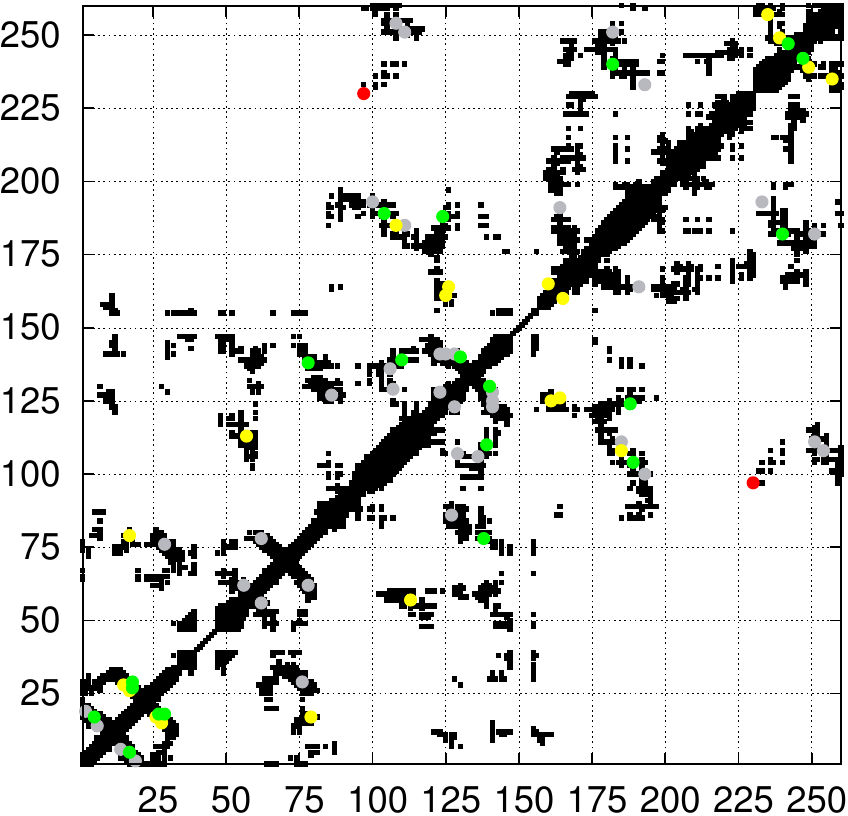} 
\caption{First $40$ predicted contacts for the PF00069 family (Protein
  Kinase domain) with Gaussian DCA, using the same settings as for
  Fig.~\ref{fig:psicov}. The left panel shows the predicted contacts
  overlaid on the PDB structure \emph{3fz1} (figure produced using the
  PyMOL software~\cite{PyMOL}); the right panel shows the predicted
  pairs overlaid on the contact map (true contacts as obtained by
  setting the threshold at 8\AA are shown in black).  In both panels,
  the color code is the following: the first $10$ predicted contacts
  are depicted in green, the next $10$ contacts in yellow, the last
  $20$ contacts in grey; the only false positive contact (occurring as
  the $24^{\textrm{th}}$ predicted pair) is shown in red.}
\label{fig:pfam69}
\end{figure}

Visual inspection of the predicted contacts does not reveal any
significant bias with respect to the residue position, nor with
respect to the sencondary or tertiary structures of the proteins. As
an example, in Fig.~\ref{fig:pfam69} we show the first 40 predicted
contacts (39 out of which are true positives) for the protein familiy
PF00069 (Protein kinase domain) using the Gaussian DCA methods with
the FN score: the pictures seem to indicate a sparse, fair sampling
across the set of all true contacts.

\begin{figure}
\centerline{\includegraphics[width=\columnwidth]{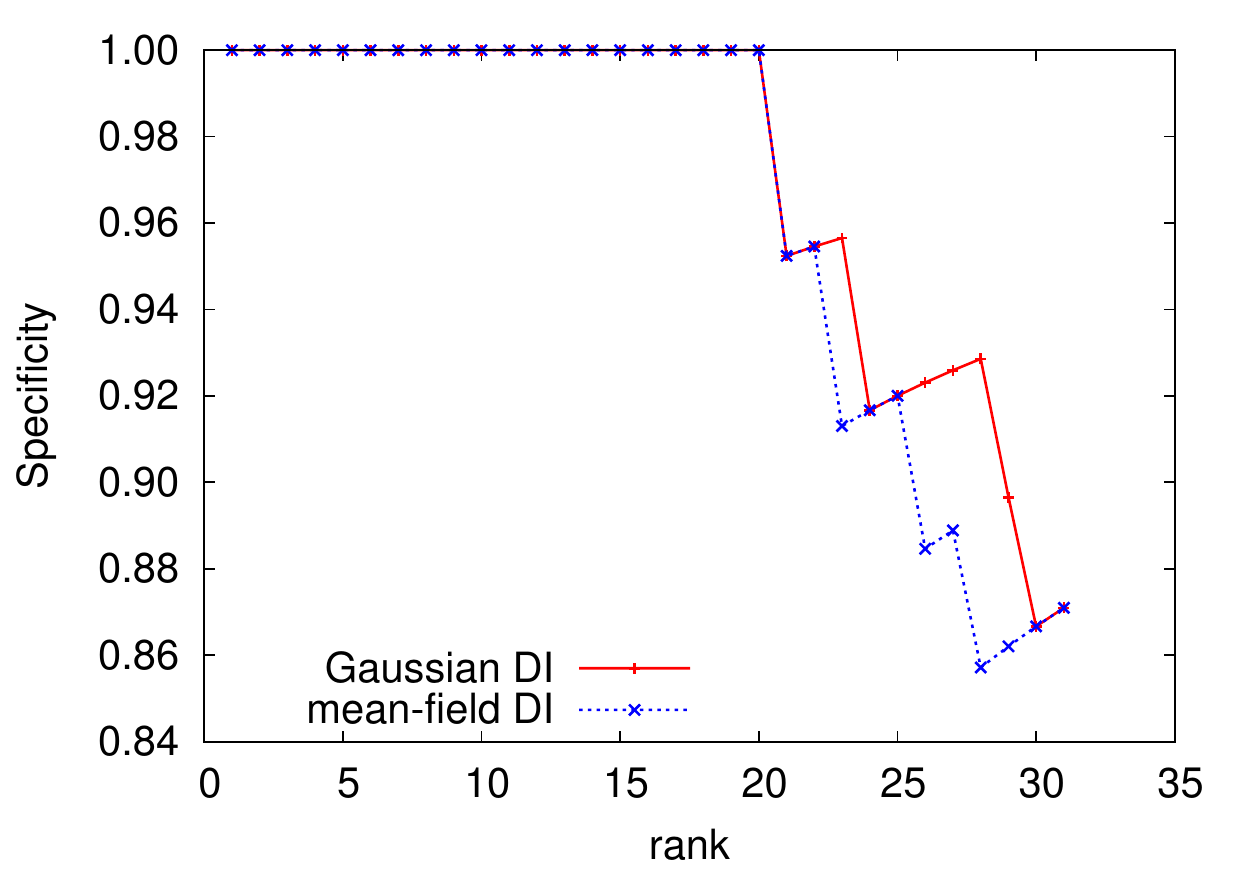}}
\caption{DI-ranking-induced mean true positive rate for predicting
\emph{inter-protein} contacts in the SK/RR complex, for both
mean-field DCA (blue curve) and multivariate Gaussian DCA (red
curve). }
\label{SK-RRrocs}
\end{figure}

Finally, we have used the SK/RR data set containing 8,998 cognate
SK/RR pairs, cf.~Methods, to predict inter-protein residue-residue
contacts. Results can be compared with those presented
in~\cite{proca}, where the original message-passing DCA was applied to
the same data-set, and 9 true contact prediction were reported before
the first false positive appeared. In Fig.~\ref{SK-RRrocs}, results
are shown for mean-field and Gaussian DCA, using the DI score: both
methods improve substantially over the message-passing scheme (20 true positive
predictions at specificity equal to one), but are highly comparable
(with a little but not significant advantage of the Gaussian scheme).
Again, we find that the improved efficiency and analytical tractability
of Gaussian DCA comes at no cost for the predictive power.

\subsection*{Predicting interactions between proteins in bacterial signal transduction}

A typical bacterium uses, on average, about 20 two-component signal
transduction systems to sense external signals, and to trigger a
specific response. In bacteria living in complex environments, the
number of different TCS may even reach 200.
While the signals and consequently the mechanisms
of signal detection vary strongly from one TCS to another, the
internal phosphotransfer mechanism from the SK to the RR, which
activates the RR, is widely conserved across bacteria:
A majority of the kinase domains of SK belong to the protein domain
family \emph{HisKA} (PF00512), all RR to family \emph{Response\_reg}
(PF00072)~\cite{pfamNew}, cf.~the Methods section.
Despite their closely related functionality, the interactions in the different
pathways have to be highly specific, to
induce the correct specific answer for each recognized external
signal.

A big fraction of SK and RR genes belonging to the same TCS pathway are co-localized in
joint operons; the identification of the correct interaction
partner is therefore trivial: such pairs are called cognate SK/RR\@. However,
about 30\% of all SK and 55\% of all
RR are so-called orphan proteins: their genes are isolated from
potential interaction partners in the genome. While a large fraction
of the RR are expected to be involved in other signal-transduction
processes like chemotaxis, for each of the SK at least one target
RR is expected to exist. It is a major challenge in systems biology
to identify these partners, and to unveil the signaling networks
acting in the bacteria. A step in this direction was taken
in~\cite{nimwegen,proca}, where co-evolutionary information extracted
from cognate pairs is used to predict, with some success, orphan
interaction partners.

An approach based on message-passing DCA~\cite{proca} was tested in
two well-studied model bacteria, namely
\emph{Caulobacter crescentus} (CC) and \emph{Bacillus subtilis} (BS),
where several orphan interactions are known experimentally~\cite{BS,
CC1, CC2}. The degree of accuracy of the method can be evinced from figure 4
of~\cite{proca}: for CC, all known interactions between DivL, PleC,
DivJ and CC\_1062 with DivK and PleD are correctly reconstructed by
the ranking obtained from the co-evolutionary scoring. Only in the case of the pair CenK-CenR,
the signal is not sufficiently strong.  For BS all the 5 orphan
kinases KinA-B-C-D-E are known to interact with the RR Spo0F, which was
clearly visible in co-evolutionary analysis in all but the KinB case.

\begin{figure}
\centerline{\includegraphics[width=\columnwidth]{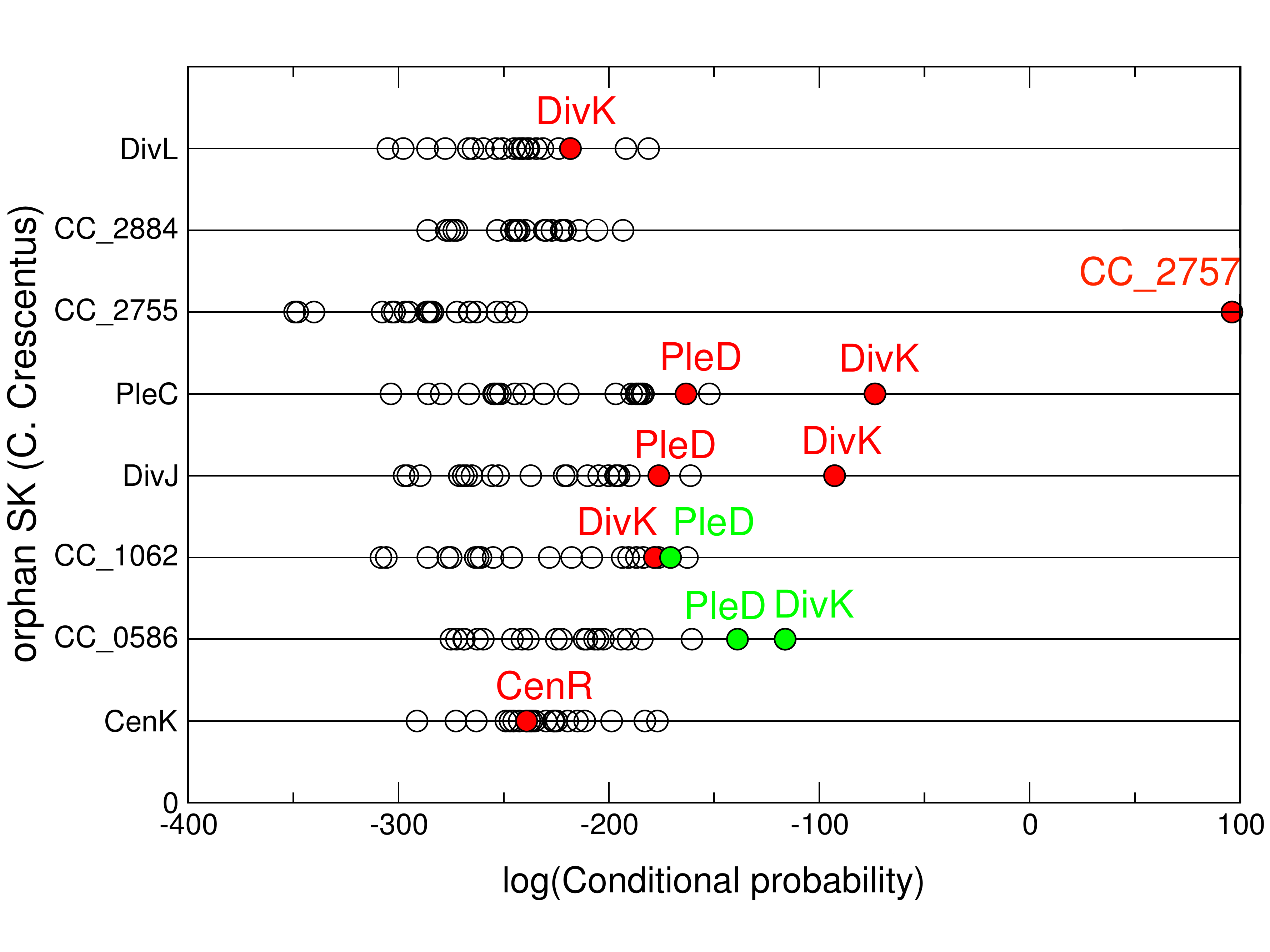}}
\caption{Partner prediction for \emph{Caulobacter crescentus} orphan
two-component proteins by the conditional probability method.
Experimentally known interaction partners~\cite{CC1,CC2} are shown
in red. Green dots correspond to partner predictions suggested
in~\cite{proca}. As for~\cite{proca}, the overall performance of the algorithm
is good, except for the prediction on CenK-CenR interaction.}
\label{CCorph}
\end{figure}

\begin{figure}
\centerline{\includegraphics[width=\columnwidth]{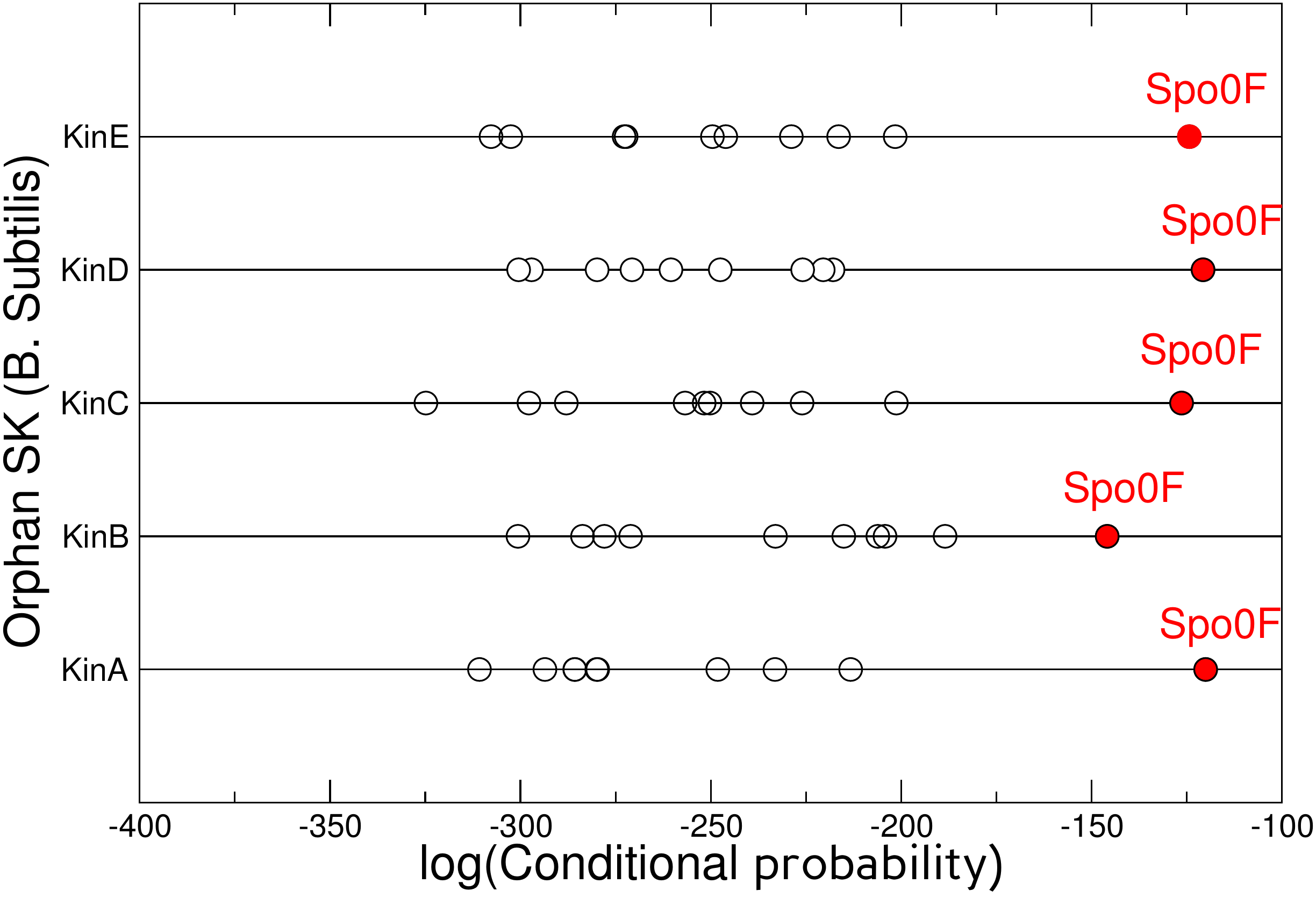}}
\caption{Partner prediction for \emph{Bacillus subtilis} orphan two-component
proteins. All 5 orphan kinases, KinA-E, are known to phosphorylate
Spo0F, which is displayed in red and is always the
maximally scoring protein in the RR set.}
\label{BSorph}
\end{figure}

The method proposed here for orphans pairing relies on the Gaussian
approximation and on the definition of the score $\Lk$,
cf.~Eq.~\ref{eq:L_score} in Methods, which equals the log-odds ratio between
the probabilities
of two orphan sequences in the interacting model (inferred from
cognate SK/RR alignments) and a non-interacting model (inferred
independently from the two MSAs of the SK and the RR families). It is
worth stressing at this point that all estimates of the likelihood
score parameters are learned only on the cognates set. Ranked by
$\Lk$, orphans interactions in CC are shown in Fig.~\ref{CCorph}.
Results are very similar to those mentioned for~\cite{proca}: known
interactions are well reproduced for orphan kinases PleC and DivJ,
while for CC\_1062 and DivL the signal for an interaction with DivK,
though present, is less clear. Finally, predictions for CC\_0586 are
identical in both studies but neither one is able to identify the
CenK-CenR interaction. Fig.~\ref{BSorph} shows predictions for orphan
interactions in BS: observed interactions between KinA, KinB, KinC,
KinD, KinE and Spo0F are manifest. This means that while predictions
in CC are slightly less accurate compared to the message-passing
strategy, predictions in BS show a greater accuracy.

\subsection*{Discussion}

In this work we have derived a multivariate Gaussian approach to
co-evolutionary analysis, whereby we cast the problem of the inference of
contacts in MSAs, as well as candidate interacting partners within two MSAs of
interacting proteins, into a simple Bayesian formalism, under the hypothesis of
normal inverse Wishart distribution of the Gaussian parameters.

The major advantage of this method is the very simple structure of the
resulting probability distribution, which allows to derive analytical
expressions for many relevant quantities (e.g.~likelihoods and posterior
probabilities). As a result, the computations performed with this model can be
very efficient, as demonstrated by the code accompanying this paper.

Furthermore, our tests indicate that the prediction accuracy of residue
contacts using the Gaussian model is comparable or superior to that achieved
using the mean-field Potts model of~\cite{Morcos2011}, or by using the PSICOV
method of~\cite{Jones} with default settings; accuracy in pairing interaction
partners is comparable to that achieved in~\cite{proca}.

The simplicity and tractability of the model also suggests further directions
for improvement. For example, the whole posterior distribution of relevant
observables such as the DI could be studied and, possibly, used to provide more
insight into the kind of predictions presented here (in particular, it could be
used to measure the confidence on the predictions).  Also, suitably designed,
more informative priors (e.g.~carrying biologically relevant information) could
further enhance the prediction power of the method, although it is not obvious
how to set a prior directly on the predicted interaction strengths, whereas
with other methods -- notably plmDCA~\cite{Aurell2013} and PSICOV~\cite{Jones}
-- this should be straightforward. Finally, we observe that the log-likelihood
score for interaction partners does not require an interaction model to be
known in advance: the interaction partners can be identified across the whole
families by optimizing the score of the joint alignment as a function of the
mapping between potentially interacting partners, thus allowing to infer both
the interacting elements and their inter-protein contacts at once.

\section*{Materials and Methods}

\subsection*{Data}

Input data is given as multiple sequence alignments of protein domains. For
the first question (inference of residue-residue contacts in protein
domains), we directly use MSAs downloaded from the Pfam
database version 27.0~\cite{pfam,pfamNew}, which are generated by aligning
successively sequences to profile hidden Markov models (HMMs)~\cite{Eddy} generated from
curated seed alignments. We have selected 50 domain families, which
were chosen according to the following criteria: \emph{(i)}~each
family contains at least 2,000 sequences, to provide sufficient
statistics for statistical inference; \emph{(ii)}~each family has at
least one member sequence with an experimentally resolved high-resolution
crystal structure available from the Protein Data Bank (PDB)~\cite{PDB}, for
assessing \emph{a posteriori} the predictive quality of the purely
sequence-based inference. The average sequence length of these 50 MSAs is
$\langle L \rangle \simeq 173$ residues, the longest sequences are those of
family PF00012 whose profile HMM contains $602$ residues. The list of included
protein domains, together with their PDB structure, is provided in
Table~\ref{table:pfam_table}.

\begin{center}
\begin{longtable}{|c|l|l|}

\hline \multicolumn{1}{|c|}{\textbf{Pfam ID}} & \multicolumn{1}{l|}{\textbf{Description}} & \multicolumn{1}{l|}{\textbf{PDB}} \\ \hline \hline
\endfirsthead

\multicolumn{3}{c} {{\bfseries \tiny{\tablename\ \thetable{} -- continued from previous page}}} \\
\hline \multicolumn{1}{|c|}{\textbf{Pfam ID}} & \multicolumn{1}{l|}{\textbf{Description}} & \multicolumn{1}{l|}{\textbf{PDB}} \\ \hline \hline
\endhead

%\multicolumn{3}{|r|}{{\tiny{Continued on next page}}} \\ \hline
\multicolumn{3}{c} {{\bfseries \tiny{\tablename\ \thetable{} -- continues on next page}}} \\
\endfoot

%\hline \hline
\endlastfoot

\small{PF00001} & \small{7 transmembrane receptor (rhodopsin family)}                      & \small{1f88, 2rh1} \\
\hline
\small{PF00004} & \small{ATPase family associated with various cellular activities (AAA)}  & \small{2p65, 1d2n} \\
\hline
\small{PF00006} & \small{ATP synthase alpha/beta family, nucleotide-binding domain}        & \small{2r9v}       \\
\hline
\small{PF00009} & \small{Elongation factor Tu GTP binding domain}                          & \small{1skq, 1xb2} \\
\hline
\small{PF00011} & \small{Hsp20/alpha crystallin family}                                    & \small{2bol}       \\
\hline
\small{PF00012} & \small{Hsp70 protein}                                                    & \small{2qxl}       \\
\hline
\small{PF00013} & \small{KH domain}                                                        & \small{1wvn}       \\
\hline
\small{PF00014} & \small{Kunitz/Bovine pancreatic trypsin inhibitor domain}                & \small{5pti}       \\
\hline
\small{PF00016} & \small{Ribulose bisphosphate carboxylase large chain, catalytic domain}  & \small{1svd}       \\
\hline
\small{PF00017} & \small{SH2 domain}                                                       & \small{1o47}       \\
\hline
\small{PF00018} & \small{SH3 domain}                                                       & \small{2hda, 1shg} \\
\hline
\small{PF00025} & \small{ADP-ribosylation factor family}                                   & \small{1fzq}       \\
\hline
\small{PF00026} & \small{Eukaryotic aspartyl protease}                                     & \small{3er5}       \\
\hline
\small{PF00027} & \small{Cyclic nucleotide-binding domain}                                 & \small{3fhi}       \\
\hline
\small{PF00028} & \small{Cadherin domain}                                                  & \small{2o72}       \\
\hline
\small{PF00032} & \small{Cytochrome b(C-terminal)/b6/petD}                                 & \small{1zrt}       \\
\hline
\small{PF00035} & \small{Double-stranded RNA binding motif}                                & \small{1o0w}       \\
\hline
\small{PF00041} & \small{Fibronectin type III domain}                                      & \small{1bqu}       \\
\hline
\small{PF00042} & \small{Globin}                                                           & \small{1cp0}       \\
\hline
\small{PF00043} & \small{Glutathione S-transferase, C-terminal domain}                     & \small{6gsu}       \\
\hline
\small{PF00044} & \small{Glyceraldehyde 3-phosphate dehydrogenase, NAD binding domain}     & \small{1crw}       \\
\hline
\small{PF00046} & \small{Homeobox domain}                                                  & \small{2vi6}       \\
\hline
\small{PF00056} & \small{Lactate/malate dehydrogenase, NAD binding domain}                 & \small{1a5z}       \\
\hline
\small{PF00059} & \small{Lectin C-type domain}                                             & \small{1lit}       \\
\hline
\small{PF00064} & \small{Neuraminidase}                                                    & \small{1a4g}       \\
\hline
\small{PF00069} & \small{Protein kinase domain}                                            & \small{3fz1}       \\
\hline
\small{PF00071} & \small{Ras family}                                                       & \small{5p21}       \\
\hline
\small{PF00072} & \small{Response regulator receiver domain}                               & \small{1nxw}       \\
\hline
\small{PF00073} & \small{Picornavirus capsid protein}                                      & \small{2r06}       \\
\hline
\small{PF00075} & \small{RNase H}                                                          & \small{1f21}       \\
\hline
\small{PF00077} & \small{Retroviral aspartyl protease}                                     & \small{1a94}       \\
\hline
\small{PF00078} & \small{Reverse transcriptase (RNA-dependent DNA polymerase)}             & \small{1dlo}       \\
\hline
\small{PF00079} & \small{Serpin (serine protease inhibitor)}                               & \small{1lj5}       \\
\hline
\small{PF00081} & \small{Iron/manganese superoxide dismutases, alpha-hairpin domain}       & \small{3bfr}       \\
\hline
\small{PF00082} & \small{Subtilase family}                                                 & \small{1p7v}       \\
\hline
\small{PF00084} & \small{Sushi domain (SCR repeat)}                                        & \small{1elv}       \\
\hline
\small{PF00085} & \small{Thioredoxin}                                                      & \small{3gnj}       \\
\hline
\small{PF00089} & \small{Trypsin}                                                          & \small{3tgi}       \\
\hline
\small{PF00091} & \small{Tubulin/FtsZ family, GTPase domain}                               & \small{2r75}       \\
\hline
\small{PF00092} & \small{Von Willebrand factor type A domain}                              & \small{1atz}       \\
\hline
\small{PF00102} & \small{Protein-tyrosine phosphatase}                                     & \small{1pty}       \\
\hline
\small{PF00104} & \small{Ligand-binding domain of nuclear hormone receptor}                & \small{1a28}       \\
\hline
\small{PF00105} & \small{Zinc finger, C4 type (two domains)}                               & \small{1gdc}       \\
\hline
\small{PF00106} & \small{Short chain dehydrogenase}                                        & \small{1a27}       \\
\hline
\small{PF00107} & \small{Zinc-binding dehydrogenase}                                       & \small{1a71}       \\
\hline
\small{PF00108} & \small{Thiolase, N-terminal domain}                                      & \small{3goa}       \\
\hline
\small{PF00109} & \small{Beta-ketoacyl synthase, N-terminal domain}                        & \small{1ox0}       \\
\hline
\small{PF00111} & \small{2Fe-2S iron-sulfur cluster binding domain}                        & \small{1a70}       \\
\hline
\small{PF00112} & \small{Papain family cysteine protease}                                  & \small{1o0e}       \\
\hline
\small{PF00113} & \small{Enolase, C-terminal TIM barrel domain}                            & \small{2al2}       \\
\hline

\caption {50 Pfam families used in the benchmarks,
         together with their associated PDB entries} \label{table:pfam_table} \\

\end{longtable}
\end{center}

Following~\cite{Jones}, we discarded the sequences in which the fraction
of gaps was larger then $0.9$. However, in~\cite{Jones}, an additional
pre-processing stage was applied, in which a target sequence is chosen
as the one for which prediction of contacts is desired, and all residue
positions in the alignment (i.e.~columns in the alignment matrix $X$) where
the target sequence alignment has gaps are removed. We did not find this
pre-processing step to improve the prediction, for either PSICOV or
our model, and therefore all results presented in this work do not
include this additional filtering.

For the second question (identification of interaction partners), we
have used the data of~\cite{proca}, thus having the possibility to
directly compare with previous results. In summary (for details
see~\cite{proca}), this data comes from 769 bacterial genomes, scanned
using HMMER2 with the Pfam 22.0 HMMs for the Sensor Kinase (SK) domain \emph{HisKA}
(PF00512) and for the Response Regulator domain \emph{Response\_reg}
(PF00072)~\cite{Finn}, resulting in 12,814 SK and 20,368 RR sequences.

A total of 8,998 SK-RR pairs are found to be cognates, i.e.~to be
coded by genes in common operons, while the rest are so-called
orphans. For statistical inference, cognates sequences are
concatenated into a single MSA, each line containing exactly one SK
and its cognate RR\@.

\subsection*{A binary representation of MSA}

The data we use are MSAs for large protein-domain families. An MSA provides a
$M\times L$-dimensional array
$A=\left(a_l^m\right)_{l=1,\ldots,L}^{m=1,\ldots,M}$: each row contains one of
the $M$ aligned homologous protein sequences of length $L$. Sequence alignments
are formed by the $Q=20$ different amino-acids, and may contain alignment gaps,
and therefore the total alphabet size is $Q+1=21$.  For simplicity, we denote
amino-acids by numbers $1,\ldots,20$, and the gap by $21$.

\begin{figure}
\centerline{\includegraphics[width=\columnwidth]{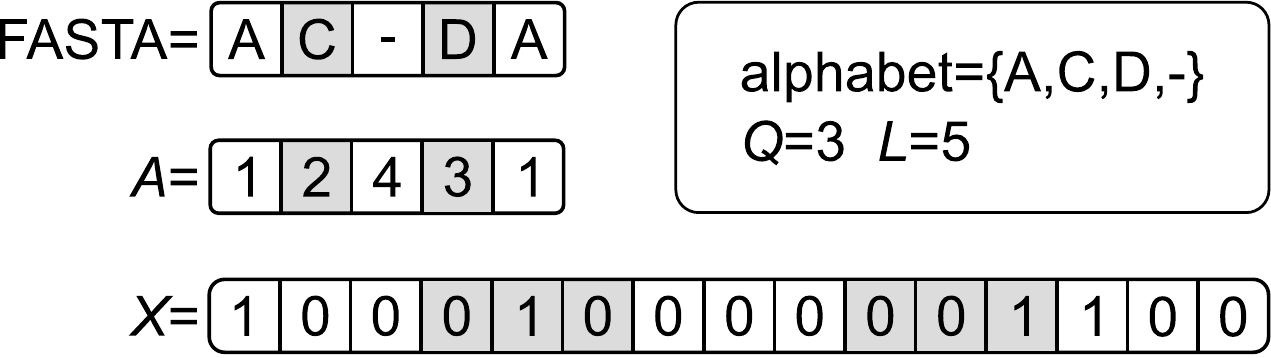}}
\caption{Illustration of the encoding of a sequence from FASTA format to its
intermediate numeric representation (matrix $A$) to its final binarized
representation (matrix $X$). For clarity, we restrict the alphabet to $Q=3$
amino-acids, $\left\{A,C,D\right\}$, plus the gap.  The alternation of
white and gray cell backgrounds helps to track the transformation (e.g.~$C
\to 2 \to 010$). Typically, MSAs of protein families are such that in every
column (i.e.~residue position) there appears a number of distinct residues
smaller than or equal to $Q=20$. Here, we did not not consider a
restriction of the alphabet to the residues actually occurring, and we used
instead the same encoding for all residues.}
\label{fig_encoding}
\end{figure}

Here we consider a modified representation, similar to that used in
\cite{Jones}, which turns out to be more practical for the
multivariate modeling we are going to propose
(cf.~Fig.~\ref{fig_encoding}).  The MSA is transformed into a $M
\times \left(Q \cdot L\right)$-dimensional array
$X=\left(x_i^m\right)_{i=1,\ldots,QL}^{m=1,\ldots,M}$ over a binary
alphabet $\{0,1\}$.  More precisely, each residue position in the
original alignment is mapped to $Q$ binary variables, each one
associated with one standard amino-acid, taking value one if the
amino-acid is present in the alignment, and zero if it is absent; the
gap is represented by $Q$ zeros (i.e.~no amino-acid is present).
Consequently, at most one of the $Q$ variables can be one for a given
residue position. For each sequence, the new variables are collected
in one row vector, i.e.~$x_{\left(l-1\right)Q+a}^m = \delta_{a,a_l^m}$
for $l=1,\ldots,L$ and $a=1,\ldots,Q$. The Kronecker symbol
$\delta_{a,b}$ equals one for $a=b$, and zero otherwise.

Denoting the row length of $X$ as $N=QL$, we introduce its empirical
mean $\overline x = \left(\overline x_i\right)_{i=1,\ldots,N}$ and the empirical
covariance matrix $C\left(X,\mu\right) = \left(C\left(X,\mu\right)_{ij}\right)_{i,j=1,\ldots,N}$ for given mean
$\mu=\left(\mu_i\right)_{i=1,\ldots,N}$:
\begin{eqnarray}
    \label{eq:bar_x}
    \overline x_i &=&
        \frac 1 M \sum_{m=1}^M x_i^m \ , \\
    \label{eq:bar_C}
    C_{ij}\left(X,\mu\right) &=&
        \frac 1 M
        \sum_{m=1}^M
            \left( x_i^m - \mu_i \right)
            \left( x_j^m - \mu_j \right).
\end{eqnarray}
The empirical covariance is thus $\overline{C} = C\left(X,\overline
x\right)$. Note that the entry $\overline x_i$, with
$i=\left(l-1\right)Q+a$, measures the fraction of proteins having
amino-acid $a \in \{1,\ldots,Q\}$ at position $l \in \{1,\ldots,L\}$.
Similarly, the entry $C_{ij}\left(X,0\right)$ of the correlation
matrix, with $i=\left(k-1\right)Q+a$ and $j=\left(l-1\right)Q+b$, is
the fraction of proteins which show simultaneously amino-acid $a$ in
position $k$ and $b$ in position $l$.

\subsubsection*{The Gaussian model}

We develop our multivariate Gaussian approach by approximating the
binary variables as real-valued variables. Even though the former are
highly structured, due to the fact that at most one amino-acid is
present in each position of each sequence, we will not enforce these
constraints on the model.  Instead, we shall rely on the fact that the
constraint is present by construction in the input data, and that as a
consequence we have, for any residue position $l$ and any two states
$a$ and $b$ with $a\ne b$:
\begin{equation}
    C_{\left(l-1\right)Q+a, \left(l-1\right)Q+b} =
        -\overline x_{\left(l-1\right)Q+a} \overline x_{\left(l-1\right)Q+b} \le 0
\end{equation}
i.e.~two different amino-acids at the same site are anti-correlated. Therefore,
we shall let the parameter inference machinery work out suitable
couplings between different amino-acid values at the same site, which
generate these observed anti-correlations.

The multivariate Gaussian model and the Bayesian inference of its
parameters are well-studied subjects in statistics, thus here we only
briefly review the main ideas behind our approach, referring
to~\cite{GelmanBook} for details.  The multivariate Gaussian
distribution is parametrized by a mean vector
$\mu=\left(\mu_i\right)_{i=1,\ldots,N}$ and a covariance matrix
$\Sigma=\left(\Sigma_{ij}\right)_{i,j=1,\ldots,N}$. Its probability
density is
\begin{equation}
P(x|\mu,\Sigma)=(2\pi)^{-\frac{N}{2}}|\Sigma|^{-\frac{1}{2}}\exp\left[ -\frac 12 (x-\mu)^T\Sigma^{-1}(x-\mu)\right],
\end{equation}
$|\Sigma|$ being the determinant of $\Sigma$, and it turns out that
the $Q\times Q$ block
\begin{equation}
    \label{eq:e}
    e_{kl}\left(a,b\right)=-\left(\Sigma^{-1}\right)_{\left(k-1\right)Q+a,\left(l-1\right)Q+b}
\end{equation}
(with $k,l\in\left\{1,\ldots,L\right\}$ and~$a,b\in\left\{1,\ldots,Q\right\}$)
plays the role of the direct interaction term in DCA between residues
$k$ and $l$.
Assuming for the moment statistical independence of the
$M$ different protein sequences in the MSA, the probability of the
data $X$ under the model (i.e.~the likelihood) reads
\begin{equation}
\label{eq:likelihood}
P\left(X|\mu,\Sigma\right)=\prod_{m=1}^M P\left(x^m|\mu,\Sigma\right)
=\left(2\pi\right)^{-\frac{NM}{2}} |\Sigma|^{-\frac{M}{2}} \exp \left[ -\frac M 2 \mathrm{tr}
        \left( \Sigma^{-1} C \left(X, \mu \right) \right)\right],
\end{equation}
%\begin{eqnarray}
%\label{eq:likelihood}
%P\left(X|\mu,\Sigma\right)
%&=& \prod_{m=1}^M \frac {
%      \exp\left[ -\frac 12 \sum_{i,j=1}^N \left(x_i^m -\mu_i\right)
%        \left(\Sigma^{-1}\right)_{ij}
%         \left(x_j^m -\mu_j\right)
%      \right] }
%         {\sqrt{\left(2\pi\right)^N \det\left(\Sigma\right)}}
%    \nonumber \\
%&=& \frac{ \exp \left[ -\frac M 2 \mathrm{tr}
%        \left( \Sigma^{-1} C \left(X, \mu \right) \right)
%    \right]}{\left[
%        \left(2 \pi\right)^N \det \left( \Sigma \right)
%    \right]^{\frac {M} 2}},
%\end{eqnarray}
with $C\left(X,\mu\right)$ given by Eq.~\ref{eq:bar_C}. %Note that each
%factor in the first line contains the model distribution for a single
%amino-acid sequence, thus it turns out that the partial correlations
%\begin{equation}
%    \label{eq:e}
%    e_{kl}\left(a,b\right):=-\left(\Sigma^{-1}\right)_{\left(k-1\right)Q+a,\left(l-1\right)Q+b}
%\end{equation}
%play the role of the direct interaction terms in DCA.

When the empirical covariance $\overline{C}$ is full rank, the
likelihood attains its maximum at $\mu=\overline x$ and $\Sigma =
\overline{C}$, which constitute the parameter estimates within the
maximum likelihood approach. However, due to the under-sampling of the
sequence space, $\overline{C}$ is typically rank deficient and this
inference method is unfeasible. To estimate proper parameters, we make
use of a Bayesian inference method, which needs the introduction of a
prior distribution over $\mu$ and $\Sigma$. The required estimate is
then computed as the mean of the resulting posterior, which is the
parameter distribution conditioned to the data. As we have already
mentioned, a convenient prior is the conjugate prior, which gives a
posterior with the same structure as the prior but identified by
different parameters accounting for the data contribution. The
conjugate prior of the multivariate Gaussian distribution is the
normal-inverse-Wishart (NIW) distribution. A NIW prior has the form
$p\left(\mu,\Sigma\right)=p\left(\mu|\Sigma\right)p\left(\Sigma\right)$, where
\begin{equation}
\label{eq:prior_mu}
p\left(\mu|\Sigma\right)
=(2\pi)^{-\frac{N}{2}}\kappa^{\frac{N}{2}}|\Sigma|^{-\frac{1}{2}}\exp\left[-\frac{\kappa}{2}\left(\mu-\eta\right)^{T}\Sigma^{-1}\left(\mu-\eta\right)\right]
\end{equation}
is a multivariate Gaussian distribution on $\mu$ with covariance
matrix $\Sigma/\kappa$ and prior mean
$\eta=\left(\eta_i\right)_{i=1,\ldots,N}$. The parameter $\kappa$ has the meaning
of number of prior measurements. The prior on $\Sigma$ is the
inverse-Wishart distribution
\begin{equation}
\label{eq:prior_S}
p\left(\Sigma\right)=\frac{1}{Z}\left|\Sigma\right|^{-\frac{\nu+N+1}{2}}\exp\left[-\frac{1}{2}\mathrm{tr}
\left(\Lambda\Sigma^{-1}\right)\right],
\end{equation}
where $Z$ is a normalizing constant:
\begin{equation}
Z=2^{\frac{\nu N}{2}}\pi^{\frac{N(N-1)}{4}}|\Lambda|^{-\frac{\nu}{2}}\prod_{n=1}^N\Gamma\left(\frac{\nu+1-n}{2}\right),
\end{equation}
$\Gamma$ being Euler's Gamma function. The parameters $\nu$ and
$\Lambda=\left(\Lambda_{ij}\right)_{i,j=1,\ldots,N}$ are the degree of freedom
and the scale matrix, respectively, shaping the inverse-Wishart
distribution.  The condition for this distribution to be integrable is
$\nu>N-1$.  The posterior $p\left(\mu,\Sigma|X\right)$, proportional to
$P\left(X|\mu,\Sigma\right)\cdot p\left(\mu,\Sigma\right)$, is still a NIW distribution, as
one can easily verify starting from Eqs.~\ref{eq:likelihood},
\ref{eq:prior_mu} and~\ref{eq:prior_S}. The posterior distribution
$p\left(\mu,\Sigma|X\right)$ is characterized by parameters $\kappa^\prime$,
$\eta^\prime$, $\nu^\prime$, and $\Lambda^\prime$ given by the
formulae
%\begin{equation}
%p\left(\mu,\Sigma|X\right)=\frac{1}{Z_{M}}\left|\Sigma\right|^{-\left(\frac{\nu_{M}+N}{2}+1\right)}\exp\left[-\frac{1}{2}\mathrm{tr}
%\left(\Lambda_{M}\Sigma^{-1}\right)-\frac{\kappa_{M}}{2}\left(\mu-\mu_{M}\right)^{T}\Sigma^{-1}\left(\mu-\mu_{M}\right)\right]
%\end{equation}
%\begin{eqnarray}
%\kappa^\prime & = & \kappa+M\\
%\eta^\prime & = & \frac{\kappa}{\kappa+M}\eta+\frac{M}{\kappa+M}\overline x\\
%\nu^\prime & = & \nu+M\\
%\Lambda^\prime & = & \Lambda+\overline{C}+\frac{\kappa M}{\kappa+M}\left(\overline x-\eta\right)\left(\overline x-\eta\right)^{T}
%\end{eqnarray}
\begin{equation}
\begin{cases}
\displaystyle{\kappa^\prime =  \kappa+M},\\
\displaystyle{\eta^\prime = \frac{\kappa}{\kappa+M}\eta+\frac{M}{\kappa+M}\overline x},\\
\displaystyle{\nu^\prime = \nu+M},\\
\displaystyle{\Lambda^\prime = \Lambda+M\overline{C}+\frac{\kappa M}{\kappa+M}\left(\overline x-\eta\right)\left(\overline x-\eta\right)^{T}}.
\end{cases}
\end{equation}
The mean values of $\mu$ and $\Sigma$ under the NIW prior are $\eta$
and $\Lambda/\left(\nu-N-1\right)$, and, similarly, their expected
values under the NIW posterior are $\eta^\prime$ and
$\Lambda^\prime/\left(\nu^\prime-N-1\right)$, respectively. Our
estimates of the mean vector and the covariance matrix, that with a
slight abuse of notation we shall still denote by $\mu$ and $\Sigma$
for the sake of simplicity, are thus
\begin{equation}
\mu=\eta^\prime=\frac{\kappa}{\kappa+M}\eta+\frac{M}{\kappa+M}\overline x
\label{eq:muNIW}
\end{equation}
and
\begin{equation}
\Sigma=\frac{\Lambda^\prime}{\nu^\prime-N-1}=\frac{\Lambda +M\overline{C}+\frac{kM}{k+M}\left(\overline x-\eta\right)^T
\left(\overline x-\eta\right)}{\nu+M-N-1}.
\label{eq:SNIW}
\end{equation}
The NIW posterior is maximum at $\mu=\eta^\prime$ and
$\Sigma=\Lambda^\prime/(\nu^\prime+N+1)$, with the consequence that
the \textit{maximum a posteriori} estimate would provide the same
estimate of $\mu$ and an estimate of $\Sigma$ that only differs
from the previous one by a scale factor.
%The corresponding parameter estimates are:
%\begin{eqnarray}
%    \mu &=& \frac{k}{k+M}\eta + \frac{M}{k+M} \overline x\\
%    \label{eq:muNIW}
%    \Sigma &=& \frac{ \Lambda +M\overline{C}+\frac{kM}{k+M}\left(\overline x-\eta\right)^T\left(\overline x-\eta\right)}{\nu+M-N-1},
%    \label{eq:SNIW}
%\end{eqnarray}
%where $k$ has the meaning of prior measurements, $\eta$ is the prior
%mean, and $\nu$ and $\Lambda$ are the degrees of freedom and the scale
%matrix for the inverse-Wishart prior over $\Sigma$, respectively.

As a first attempt of protein contact prediction by means of the
present model, we choose $\eta$ and $\Lambda$ to be as uninformative as
possible. In particular, since $U=\Lambda/\left(\nu-N-1\right)$ is the
prior estimate of $\Sigma$, it is natural to set
$\eta=\left(\eta_i\right)_{i=1,\dots,N}$ and $U=\left(U_{ij}\right)_{i,j=1,\ldots,N}$ to the
mean and the covariance matrix of uniformly distributed samples.
Therefore, we set $\eta_i=1/\left(Q+1\right)$ for any $i$, and $U$ to a
block-matrix composed of $L \times L$ blocks of size $Q\times Q$ each,
where the out-of-diagonal blocks are uniformly $0$:
\begin{equation}
    U_{\left(k-1\right)Q+a,\left(l-1\right)Q+b} =
    \frac {\delta \left(k,l\right)} {Q+1}
    \left(
        \delta \left(a,b\right)
        -\frac 1 {Q+1}
    \right),
\end{equation}
where $k,l\in\left\{1,\ldots,L\right\}$ and~$a,b\in\left\{1,\ldots,Q\right\}$,
and $\delta$ is the Kronecker's symbol. Moreover, we choose
$\nu=N+\kappa+1$ in order to reconcile Eq.~\ref{eq:SNIW} with the
pseudo-count-corrected covariance matrix of~\cite{Morcos2011} with
pseudo-count parameter $\lambda$.  Indeed, identifying $\lambda$ with
$\kappa/\left(\kappa+M\right)$, this instance allows us to recast
the estimation of $\Sigma$ as
\begin{equation}
    \label{eq:MAP_Sigma}
    \Sigma = \lambda U + \left(1 - \lambda\right) \overline{C} +
             \lambda \left(1 - \lambda\right)
             \left(\overline x - \eta\right)^T
             \left(\overline x - \eta\right)
\end{equation}
and $J=\Sigma^{-1}$ becomes the same as in the mean-field Potts model.
Manifestly from here, the effect of the prior is enhanced by values of
$\lambda$ close to 1 while it is negligible when $\lambda$ approaches
0. Interestingly, the Gaussian framework provides an interpretation of
the pseudo-count correction in terms of a prior distribution, which
may allow improving the inference issue by exploiting more informative
prior choices.

\subsubsection*{Reweighted frequency counts\label{subs:reweighting}}

The approach outlined in the above sections assumes that the rows of
the MSA matrix $X$, i.e.~the different protein sequences, form an
independently and identically distributed (i.i.d.) sample, drawn from
the model distribution, cf.~Eq.~\ref{eq:likelihood}. For biological
sequence data this is not true: there are strong sampling biases due
to phylogenetic relations between species, due to the sequencing of
different strains of the same species, and due to a non-random
selection of sequenced species. The sampling is therefore clustered in
sequence space, thereby introducing spurious non-functional
correlations, whereas other viable parts of sequence space (in the
sense of sequences which would fall into the same protein family) are
statistically underrepresented. To partially remove this sampling
bias, we use the same re-weighting scheme used in the PSICOV version
1.11 code~\cite{Jones} (which is the same as that used
in~\cite{weigt,Morcos2011}, with an additional pre-processing pass to
estimate a value for the similarity threshold; see the Supporting
Information Section for details). The procedure can be seen as
generalization of the elimination of repeated sequences.

\subsubsection*{Computing the ranking score\label{subs:direct}}

Contact prediction using DCA relies on ranking pairs of residue
positions $1\leq k<l\leq L$ according to their direct interaction
strength. As mentioned before, two positions interact via a $Q\times
Q$ matrix $e_{kl}$ given by Eq.~\ref{eq:e}. To compare
two position pairs $kl$ and $k^{\prime}l^{\prime}$, we need to map
these matrices to a single scalar quantity. We have tested two
different transformations: the first one, following~\cite{weigt},
is the so-called direct information (DI), which measures the
mutual information induced only by the direct coupling $e_{kl}$
between two positions $k$ and $l$ (for a more precise definition see
Supporting Information  Section); the second one, following~\cite{Aurell2013},
is the Frobenius norm (FN) of the sub-matrix obtained by \emph{(i)}
changing the gauge of the interaction such that the sum of each row and
column is zero, and \emph{(ii)} removing the row and column corresponding
to the gap symbol. In our empirical tests (cf.~Fig.~\ref{fig:tpr}), the
FN score can reach a better overall accuracy in residues contacts
prediction; the DI score, however, also achieves good results, is
gauge-invariant, and has a clear interpretation in terms of the
underlying model: it is therefore a useful indicator to compare the
Gaussian model with the mean-field approximation to the discrete model.
In the multivariate Gaussian setting, the DI can be calculated explicitly, as
shown in the Supporting Information Section, thus resulting in a
gain in computation time as compared to the mean-field DCA
in~\cite{Morcos2011}, while achieving similar or better performance
(cf.~Fig.~\ref{fig:tpr}).

We found empirically that both the DI and the FN scores produce
slightly better results in the residue contact prediction tests when
adjusted via average-product-correction (APC), as described
in~\cite{dunn2008mutual}.

\subsection*{Summary of the residue contact prediction steps}

To summarize the previous sections, here we list the steps which
are taken in order to get from a MSA to the contact prediction:

\begin{itemize}
    \item clean the MSA by removing inserts and keeping only matched amino acids and deletions;
    \item remove the sequences for which 90\% or more of the entries are gaps;
    \item assign a weight to each sequence, and compute the reweighted
      frequency counts $\overline{C}$ and $\overline x$ (see
      Eqs.~\ref{eq:bar_x} and~\ref{eq:bar_C}, and Suporting File S1);
%    \item compute the posterior correlation matrix $\Sigma$ from Eq.~\ref{eq:MAP_Sigma};
    \item estimate the correlation matrix $\Sigma$ by means of Eq.~\ref{eq:MAP_Sigma};
    \item compute $\Sigma^{-1}$, and divide it in $Q\times Q$ blocks $e_{kl}$ (see Eq.~\ref{eq:e});
    \item for each pair $1 \leq k, l \leq L$, compute a score (DI or FN) from $e_{kl}$, thus obtaining
          an $L\times L$ symmetric matrix $S$ (with zero diagonal);
    \item apply APC to the score matrix (i.e.~subtract to each entry $S_{kl}$ the product
          of the average score over $k$ and the average score over $l$, divided by the overall
          score average -- the averages are computed excluding the diagonal), and obtain an adjusted
          score matrix $S_{kl}^\textrm{APC}$;
    \item rank all pairs $1 \leq k < l \leq L$, with $l - k > 4$, in descending order according
          to $S_{kl}^\textrm{APC}$.
\end{itemize}

\subsection*{A log-likelihood score for protein-protein interaction}

In~\cite{proca}, DCA has been used to predict RR interaction partners
for orphan SK proteins in bacterial TCS, and to detect crosstalk
between different cognate SK/RR pairs. Relying on the improved
efficiency of the multivariate Gaussian approach presented here, we
can introduce a much clearer but similarly performing definition of a
protein-protein interaction score.

This score is based on the existence of a large set of known
interaction partners: we collect them in a unified MSA, in
which each row contains the concatenation of two interacting
protein sequences, and we encode them in a matrix denoted by
$X_{\SKRR}$. The encoded MSAs restricted to each of the
single protein families are denoted by $X_{\SK}$ and $X_{\RR}$. We
estimate model parameters $\Sigma_{A}$ and $\mu_{A}$ for each of the
three alignments $X_{A}$, with $A \in \{\SK,\RR,\SKRR\}$. Whereas the
parameters for the two alignments of single protein families describe
the \emph{intra}-domain co-evolution inside each domain, the parameter
matrix $\Sigma_{\SKRR}$, obtained from the joint MSA, also models the
\emph{inter}-protein co-evolution.

In order to decide if two new sequences $x_{\SK}$ and $x_{\RR}$
interact, we first introduce the sequence $x_{\SKRR}$ as the
(horizontal) concatenation of $x_{\SK}$ with $x_{\RR}$. Next we define
a log-odds ratio comparing the probability of these sequences
under the joint SKRR-model with the one under the separate models for
SK and RR, i.e.~we calculate
\begin{eqnarray}
    \label{eq:L_score}
    \Lk \left(x_{\SK},x_{\RR}\right)
        &=& \log \frac{P\left(x_{\SKRR}|\Sigma_{\SKRR},\mu_{\SKRR}\right)}{
            P\left(x_{\SK}|\Sigma_{\SK},\mu_{\SK}\right)
            P\left(x_{\RR}|\Sigma_{\RR},\mu_{\RR}\right)}
    \nonumber \\
        &=& c-\frac 12 \left(x_{\SKRR} -\mu_{\SKRR}\right)^t
             \Sigma_{\SKRR}^{-1} \left(x_{\SKRR} -\mu_{\SKRR}\right)
    \nonumber \\
        && +\frac 12 \left(x_{\SK} -\mu_{\SK}\right)^t
             \Sigma_{\SK}^{-1} \left(x_{\SK} -\mu_{\SK}\right)
    \nonumber \\
        && +\frac 12 \left(x_{\RR} -\mu_{\RR}\right)^t
             \Sigma_{\RR}^{-1} \left(x_{\RR} -\mu_{\RR}\right)
\end{eqnarray}
with $c$ being a constant (i.e.~not depending on the
sequence $x_{\SK},x_{\RR}$) coming from the normalization of the
multivariate Gaussians. Intuitively, this score measures to what extent
the two sequences are coherent with the model of interacting SK/RR
sequences, as compared to a model which assumes them to be just two
arbitrary (and thus typically not interacting) SK and RR sequences.
In mathematical terms, it can also be seen as the log-odds ratio
between the conditional probability of $x_{\SK}$ knowing $x_{\RR}$,
and the unconditioned probability of $x_{\SK}$.

\section*{Acknowledgments}

CF acknowledges funding from the People Programme (Marie Curie Actions) of the
European Union's Seventh Framework Programme FP7/2007-2013/ under REA grant
agreement n°~290038.

CB and RZ acknowledge the European Research Council for grant n°~267915.

\section*{Supporting Informations}

\subsection*{Direct Information computation in the Gaussian model}

In order to implement DCA, we aim at quantifying the effect of the
interaction between each pair of residues. The idea is to compare a
system with only two interacting residues with the non--interacting
corresponding scene. Single--site marginals are preserved in both
cases while the interaction term is encoded in the matrix
$J=\Sigma^{-1}$ as derived in the Main Text. Indeed, the interaction
between residues $l$ and $l^\prime$ is described by $e_{l
  l^\prime}:=-\hJ_{ll^\prime}$, denoting with $\hJ_{ll^\prime}$ the
$Q\times Q$ block of $J$ corresponding to residues $l$ and $l^\prime$,
which is the $Q\times Q$ matrix with entries
$\left(J\right)_{nn^\prime}$ such that $n=l \mod Q$ and
$n^\prime=l^\prime \mod Q$.

The non--interacting case is easily approached by repeating the
Bayesian analysis described in the Main Text independently for each
residue $l$ and provides $\Pind_{l}\left(x\right)$, where now $x$ is
the $Q$-state vector associated to position $l$. As a result,
$\Pind_l$ is a Gaussian distribution with the blocks corresponding to
$l$ of $\mu$ and $\Sigma$, given in the Main Text respectively, as
mean and covariance.  The interacting instance between $l$ and
$l^\prime$ is instead characterized by the Gaussian distribution
$\Pdir_{ll^\prime}\left(x,x^\prime\right)$ with interaction
$\hJ_{ll^\prime}$ and single--site marginals $\Pind_{l}\left(x\right)$
and $\Pind_{l^\prime}\left(x^\prime\right)$. Notice that
$\Pdir_{ll^\prime}\left(x,x^\prime\right)$ reduces to
$\Pind_{l}\otimes\Pind_{l^\prime}\left(x,x^\prime\right):=\Pind_{l}\left(x\right)\Pind_{l^\prime}\left(x^\prime\right)$
when $\hJ_{ll^\prime}=0$.  In order to measure the strength of
$\hJ_{ll^\prime}$ we then define the direct information
$DI_{ll^\prime}$ between sites $l$ and $l^\prime$ as the
Kullback--Leibler divergence between $\Pdir_{ll^\prime}$ and
$\Pind_{l}\otimes \Pind_{l^\prime}$:
\begin{equation}
DI_{ll^\prime}:=KL\left( P_{ll^\prime}^{\mathrm{dir}} || \Pind_{l}\otimes \Pind_{l^\prime}\right).
\label{eq:DI}
\end{equation}
We have $DI_{ll^\prime}\ge 0$ and $DI_{ll^\prime}=0$ if $\hJ_{ll^\prime}=0$.

We stress that the expression of the direct information is
gauge--invariant, in the sense that it is independent of the
a.a.\ index omitted in the model construction.  Moreover, even though
the matrix $J$ is the same as in the mean--field approximation of the
Potts model~\cite{Morcos2011, noiplosone2011}, $DI_{ll^\prime}$ is
different.  

Here we show how to compute the direct information $DI_{ll^\prime}$
between residues $l$ and $l^\prime$. We denote with $\hat\mu_l$ the
mean associated to position $l$, which is the $Q$-vector with entries
$\left(\mu\right)_n$ such that $r\left(n\right)=l$. Similarly to $\hat
J_{ll^\prime}$, $\hat\Sigma_{ll^\prime}$ represent the $Q\times Q$
block of $\Sigma$ with entries $\left(\Sigma\right)_{nn^\prime}$ where
$r\left(n\right)=l$ and $r\left(n^\prime\right)=l^\prime$. The
Gaussian distribution $P^{\text{ind}}_l$ is characterized by mean
$\hat\mu_l$ and covariance $\hat\Sigma_{ll}$. The Gaussian
distribution $P^{\text{dir}}_{ll^\prime}$ has marginals
$P^{\text{ind}}_l$ and $P^{\text{ind}}_{l^\prime}$ and retains $\hat
J_{ll^\prime}$ to describe the interaction between the two
residues. Thus, writing its mean as $\left(\alpha,\beta\right)$ with
$\alpha,\beta\in\mathbb{R}^Q$ and its interaction (or precision)
matrix as
\begin{equation}
\left( \begin{array}{cc}
H & \hat J_{ll^\prime} \\
\hat J_{l^\prime l} & K
\end{array} \right)
\end{equation}
with $H,K\in\mathbb{R}^{Q\times Q}$ symmetric positive definite, the
marginalization constraints impose that $\alpha=\hat\mu_l$ and
$\beta=\hat\mu_{l^\prime}$ and that the diagonal block of
\begin{equation}
\left( \begin{array}{cc}
H & \hat J_{ll^\prime} \\
\hat J_{l^\prime l} & K
\end{array} \right)^{-1}
\end{equation}
equal $\hat\Sigma_{ll}$ and $\hat\Sigma_{l^\prime l^\prime}$
respectively. Exploiting the formula for the block-wise inversion of
block matrices, the conditions on $H$ and $K$ can be explicitly stated as 
\begin{equation}
\begin{cases}
H-\hat J_{ll^\prime}K^{-1}\hat J_{l^\prime l}=\hat\Sigma_{ll}^{-1};\\
K-\hat J_{l^\prime l}H^{-1}\hat J_{ll^\prime}=\hat\Sigma_{l^\prime l^\prime}^{-1}.
\end{cases}
\label{eq:HK}
\end{equation}

The direct information $DI_{ll^\prime}$ is the Kullback--Leibler divergence
\begin{equation}
DI_{ll^\prime}=KL\left(P^{\text{dir}}_{ll^\prime}||P^{\text{ind}}_l\otimes
P^{\text{ind}}_{l^\prime}\right):=\int_{\mathbb{R}^{Q}}dx\int_{\mathbb{R}^{Q}}dx^\prime
\ln
\biggl(\frac{P^{\text{dir}}_{ll^\prime}\left(x,x^\prime\right)}{P^{\text{ind}}_l\left(x\right)P^{\text{ind}}_{l^\prime}\left(x^\prime\right)}\biggr)P^{\text{dir}}_{ll^\prime}\left(x,x^\prime\right).
\end{equation}
Simple algebra shows that
\begin{equation}
DI_{ll^\prime}=\frac{1}{2}\biggl[\ln\det
\left( \begin{array}{cc}
H & \hat J_{ll^\prime} \\
\hat J_{l^\prime l} & K
\end{array} \right)
+\ln\det\hat\Sigma_{ll}+\ln\det\hat\Sigma_{l^\prime l^\prime}\biggr].
\end{equation}
Recalling that $\hat \Sigma_{ll}$ is a symmetric positive definite
matrix for any $l$ is of some help in order to explicit
$DI_{ll^\prime}$. Indeed, this fact tells us that $\hat \Sigma_{ll}$
admits the Cholesky decomposition $\hat \Sigma_{ll}=S_l S_l^T$ with
invertible Cholesky factor $S_l$.  Let us then introduce the matrices
$T_{ll^\prime}:=S_l^T\hat J_{ll^\prime} S_{l^\prime}$, $X:=S_l^T H
S_l$ and $Y:=S_{l^\prime}^T K S_{l^\prime}$. We have that
$T_{ll^\prime}^T=T_{l^\prime l}$ as a consequence of the relation
$\hat J_{l^\prime l}=\hat J_{ll^\prime}^T$ due to the symmetry of $J$.
With these definitions we can recast $DI_{ll^\prime}$ as
\begin{eqnarray}
\nonumber
2DI_{ll^\prime}&=&\ln\det
\left( \begin{array}{cc}
H & \hat J_{ll^\prime} \\
\hat J_{l^\prime l} & K
\end{array} \right)+
\ln\det
\left( \begin{array}{cc}
S_l S_l^T & 0 \\
0 & S_{l^\prime} S_{l^\prime}^T
\end{array} \right)\\
\nonumber
&=&\ln\det
\left( \begin{array}{cc}
H & \hat J_{ll^\prime} \\
\hat J_{l^\prime l} & K
\end{array} \right)+
\ln\det
\left( \begin{array}{cc}
S_l & 0 \\
0 & S_{l^\prime} 
\end{array} \right)+
\ln\det
\left( \begin{array}{cc}
S_l^T & 0 \\
0 & S_{l^\prime}^T
\end{array} \right)\\
&=&\ln\det
\left( \begin{array}{cc}
S_l^T & 0 \\
0 & S_{l^\prime}^T
\end{array} \right)\cdot
\left( \begin{array}{cc}
H & \hat J_{ll^\prime} \\
\hat J_{l^\prime l} & K
\end{array} \right)\cdot
\left( \begin{array}{cc}
S_l^T & 0 \\
0 & S_{l^\prime}^T
\end{array} \right)=
\ln\det\left( \begin{array}{cc}
X & T_{ll^\prime} \\
T_{ll^\prime}^T & Y
\end{array} \right).
\end{eqnarray}
In addition, starting from eq.~\ref{eq:HK}, we can write down
corresponding equations for $X$ and $Y$:
\begin{equation}
\begin{cases}
X=I+T_{ll^\prime}Y^{-1}T_{ll^\prime}^T;\\
Y=I+T_{ll^\prime}^T X^{-1}T_{ll^\prime},
\end{cases}
\label{eq:XY}
\end{equation}
being $I$ the identity $Q\times Q$-matrix. Notice that $X$ and $Y$
must constitute the positive definite solution of this
problem. Interestingly, the latter of these equations gives
\begin{equation}
\left( \begin{array}{cc}
X & T_{ll^\prime} \\
T_{ll^\prime}^T & Y
\end{array} \right)=
\left( \begin{array}{cc}
X & 0 \\
T_{ll^\prime}^T & I
\end{array} \right)\cdot
\left( \begin{array}{cc}
I & X^{-1}T_{ll^\prime} \\
0 & I
\end{array} \right).
\end{equation}
Then, the property of block matrices 
\begin{equation}
\det
\left( \begin{array}{cc}
A & C \\
0 & B
\end{array} \right)=
\det
\left( \begin{array}{cc}
A & 0 \\
C & B
\end{array} \right)=\det A \det B
\end{equation}
provides the formula 
\begin{equation}
DI_{ll^\prime}=\frac{1}{2}\ln\det X.
\end{equation}

As far as the solution of eq.~\ref{eq:XY} is concerned, let us observe
that $X^{-1}T_{ll^\prime}=T_{ll^\prime}Y^{-1}$, as one recognizes
multiplying the first equation by $X^{-1} T_{ll^\prime}$ on the left
and the second one by $T_{ll^\prime}Y^{-1}$ on the right. The
consequence of this identity is that the matrix $X$ satisfies the
relation $X^2-X-T_{ll^\prime}T_{ll^\prime}^T=0$, which is equivalent
to the first of eq.~\ref{eq:XY} after the substitution of
$T_{ll^\prime}Y^{-1}$ with $X^{-1}T_{ll^\prime}$. The matrix
$T_{ll^\prime}T_{ll^\prime}^T$ is symmetric positive semi-definite and
denoting with $t_{ll^\prime}^1\le t_{ll^\prime}^2\le\cdots\le
t_{ll^\prime}^Q$ its eigenvalues, not necessarily distinct, we have
that $X$ has eigenvalues
\begin{equation}
\frac{1+\sqrt{\displaystyle{1+4 t_{ll^\prime}^1}}}{2}\le\frac{1+\sqrt{\displaystyle{1+4 t_{ll^\prime}^2}}}{2}\le\cdots\le\frac{1+\sqrt{1+4 t_{ll^\prime}^Q}}{2}.
\end{equation}
The fact that $X$ is positive definite has been exploited here for
determining its spectrum. As the final result we get
\begin{equation}
DI_{ll^\prime}=\frac{1}{2}\sum_{q=1}^Q\ln\biggl(\frac{1+\sqrt{1+4t_{ll^\prime}^q}}{2}\biggr).
\end{equation}

\subsection*{Reweighting scheme}

We used the same reweighting scheme used in the PSICOV version 1.11
code~\cite{Jones}, to compensate for the sampling bias introduced by
phylogenetic relations between species. We report the details of the
computations here for convenience.  

Weights are computed in two steps: a pre-processing step which is used
to compute a similarity threshold $r$, and a weight-computation step
which is the same as that used in~\cite{weigt} and uses $r$ as a
parameter.

The similarity threshold $r$ is defined as being inversely
proportional to the average sequence identity, i.e.~the average, over
all pairs of sequences, of the fraction of identical amino-acids in
corresponding residues of two sequences. The constant of
proportionality is chosen as $0.32 \cdot 0.38 = 0.1216$, which gives
good overall results. As a further refinement, $r$ is clamped such
that its value cannot exceed $0.5$.

The threshold $r$ is then used to define neighborhoods around each
sequence: only sequences with less than $rL$ identical amino-acids are
considered to carry independent information, and so for each protein
sequence $a^m=\left(a^m_1,\ldots,a^m_L\right)$, $m=1,\ldots,M$, in the MSA we
count the number $n^m$ of sequences with at least $rL$ identical
amino-acids (including $a^m$ itself into this count), and we re-weight
the influence of the sequence by the factor $w^m=1/n^m$.  This leads
to a redefinition of the empirical means and covariances (see
eqs.~1 and~2 in the Main Text), for $1\leq i,j,\leq N$:
\begin{eqnarray}
    \bar{x}_i &=&
        \frac 1 \Meff
        \sum_{m=1}^M w^m x^m_i \\
    \bar{C}_{ij} &=&
        \frac 1 \Meff
        \sum_{m=1}^M
            w^m
            \left( x^m_i - \bar{x}_i \right)
            \left( x^m_j - \bar{x}_j \right),
\end{eqnarray}
where $\Meff=\sum_{m=1}^M w^m$ is a normalization factor, which can be
understood as the effective number of independent sequences. These
re-weighted empirical means are used for estimating the model
parameters (see eqs.~11 and~12 in the Main Text).
%\ref{eq:muNIW} and~\ref{eq:MAP_Sigma}.

\bibliography{gaussDCA}

\newpage

%\section*{Tables}

\end{document}